\documentclass[11pt]{article}
\DeclareMathAlphabet{\scr}{U}{rsfs}{m}{n}
\usepackage{latexsym}
\usepackage{epsfig}
\usepackage[mathscr]{eucal}
\usepackage{amsfonts, amscd, amsmath, amssymb}
\usepackage{tikz}	
\usepackage{xcolor}
\usepackage{cite}
\usepackage{mdframed}
\usepackage{graphicx}
\usepackage[footnotesize]{caption}
\usepackage{tabularx}
\usepackage{longtable}
\usepackage{hyperref}
\usepackage{dcolumn}
\usepackage{bm}
\usepackage{wasysym}
\usepackage[merge,numbers,compress]{natbib} 
\usepackage[affil-it]{authblk}

\hypersetup{
	colorlinks=true,        
	linkcolor=blue,          
	citecolor=red,        
	filecolor=magenta,      
	urlcolor=cyan           
}

\newcommand{\newc}{\newcommand}
\newc{\EW}{electroweak\;}
\newc{\DM}{dark matter\;}
\newc{\SM}{Standard Model\;}
\newc{\KK}{Kaluza-Klein\;}
\newc{\ff}{fragmentation function\;}
\newc{\be}{\begin{equation}}
\newc{\ee}{\end{equation}}
\newc{\bi}{\begin{itemize}}
\newc{\ei}{\end{itemize}}
\newc{\benu}{\begin{enumerate}}
\newc{\eenu}{\end{enumerate}}
\newc{\bc}{\begin{center}}
\newc{\ec}{\end{center}}
\newc{\bfig}{\begin{figure}}
\newc{\efig}{\end{figure}}
\newc{\neutone}{\tilde{\chi}^0_1}

\begin{document}
 \title{\hfill ~\\[-30mm]
\phantom{h} \hfill\mbox{\small TTK--16--39} 
\\[1cm]
\vspace{13mm}   \textbf{Constraints on leptophilic dark matter\\ from the AMS-02 experiment}}

\author[1]{L. Ali Cavasonza \thanks{E-mail: \texttt{cavasonza@physik.rwth-aachen.de}}}
\author[1]{H. Gast }
\author[2]{M. Kr\"amer }
\author[3]{M. Pellen }
\author[1]{S. Schael }

\affil[1]{I.~Physikalisches Institut, RWTH Aachen University, Sommerfeldstr.~14, 52074 Aachen, Germany}
\affil[2]{Institute for Theoretical Particle Physics and Cosmology, RWTH Aachen University, 52074 Aachen, Germany}
\affil[3]{Universit\"at W\"urzburg, Institut f\"ur Theoretische Physik und Astrophysik, D-97074 W\"urzburg, Germany }

\date{\today}
\maketitle

\begin{abstract}

The annihilation of dark matter particles in the Galactic halo of the Milky Way may lead to cosmic ray signatures that can be probed by 
the AMS-02 experiment, which has measured the composition and fluxes of charged cosmic rays with unprecedented precision.
Given the absence of characteristic spectral features in the electron and positron fluxes measured by AMS-02, we derive upper limits 
on the dark matter annihilation cross section for leptophilic dark matter models. Our limits are based on a new background model that describes all 
recent measurements of the energy spectra of cosmic-ray positrons and electrons. For thermal dark matter relics, we can exclude dark matter masses 
below about 100~GeV. We include the radiation of electroweak gauge bosons in the dark matter annihilation process and compute the antiproton 
signal that can be expected within leptophilic dark matter models. 

\end{abstract}

\newpage

\section{Introduction} \label{sec:intro}

The AMS-02 Collaboration has published the most precise,
separate measurements of the fluxes $\Phi_{\pm}$ of cosmic-ray
electrons and positrons to date~\cite{AMS14_2}. The data cover
particle energies $E$ up to 500\,GeV for positrons and 700\,GeV for
electrons, respectively. These measurements are inconsistent with pure secondary production.
This observation is among the most intriguing in cosmic ray physics, and 
various models have been proposed in the literature to explain the AMS-02 data. 
Most models invoke either exotic new physics like annihilations of dark matter 
particles, \emph{e.g.}~\cite{Feng2014,Cholis:2013psa,DiMauro:2015jxa}, or new astrophysical sources like
pulsars and their wind nebulae, \emph{e.g.}~\cite{Cholis:2013psa,Linden2013, Kohri2016}, to
explain the apparent excess of positrons. 

Advocating a pure dark matter origin for the large amount of 
positrons observed in cosmic rays at high energies would require a 
rather contrived scenario \cite{Cholis:2013psa,Boudaud:2014dta,
CiKaRaSt08,DoMaBrDeSa08,Bergstrom:2009fa,Finkbeiner:2010sm,Yuan:2013eja,
Jin:2013nta,Bertone:2008xr}, as very large cross sections 
are necessary to accommodate  the measured fluxes.
For this reason hybrid models are introduced in which an unspecified astrophysical
background creates a smooth spectrum of positrons (and electrons), 
while dark matter could be responsible for small additional spectral
features on top of the smooth background curve,
\emph{e.g.}~\cite{Bergstrom:2013jra,Ibarra2014}. 
Previous studies \emph{e.g.}~\cite{Bergstrom:2013jra} employed the 
positron fraction ($e^+/(e^++e^-)$) data from AMS and the simplified
phenomenological model introduced by the AMS Collaboration~\cite{AMS13}.
The overall normalisation of the electron flux,
needed to compare model predictions to data, was then often derived
from the measurement of the $e^++e^-$ flux by the Fermi-LAT
detector. 
This approach, however, is problematic for several reasons: 
First, meanwhile the AMS $e^++e^-$ data have been published.
While the AMS data sets are self-consistent, \textit{i.e.} one can derive 
both the positron fraction and the $e^++e^-$ flux from the individual fluxes
and conversely, the AMS $e^++e^-$ flux is not consistent with the Fermi-LAT 
$e^++e^-$ flux within the quoted experimental uncertainties. 
Therefore these two data sets cannot be combined in a fit in a trivial way.
Second, the uncertainties on the energy scales have to be taken into account
when combining data sets from different experiments.
Third, the data from AMS and from Fermi-LAT were taken at different times. 
The model introduced in~\cite{AMS13} to describe the positron fraction 
does not contain any time dependent parameters. The data from previous experiments
and the data from AMS~\cite{Schael2016} clearly show a time variation of the positron fraction
at energies below $\sim$20 GeV.
To avoid these issues we use only the published data from AMS 
and we introduce a new phenomenological model that properly describes the energy 
and time dependence of both cosmic-ray positrons and electrons.
We determine the best-fit values of the model parameters, and discuss how the model can be applied for searches of 
spectral signatures of exotic processes using the AMS data. 
We consider a generic leptophilic dark matter model and derive upper limits 
on the dark matter annihilation cross section from the absence of characteristic spectral features in the electron and positron fluxes measured by AMS-02. We also assess the impact 
of the uncertainty from cosmic ray propagation and the dark matter halo model on the cross section limits. Electroweak gauge boson radiation in the dark matter annihilation process will lead to 
a flux of Standard Model particles from the decay and hadronisation of the electroweak gauge bosons, including in particular antiprotons. We show that the antiproton 
flux provides a sensitive and complementary probe of dark matter models, even within the leptophilic scenario we consider. 

The article is structured as follows: In section~\ref{sec:dm} we discuss the class of dark matter models we consider and
describe how we obtain the electron and positron fluxes due to dark matter annihilation in the Galactic halo. The new background model 
is introduced in section~\ref{sec:bkg} where we also determine the best-fit values of the model parameters. The calculation 
of the limits on the dark matter annihilation cross section from AMS-02 data is presented in section~\ref{sec:UL}.  Antiproton fluxes are generated from  
the radiation of electroweak gauge bosons and may lead to complementary constraints on the dark matter model, as discussed in 
section~\ref{sec:pbar}. We summarise and conclude in section~\ref{sec:conclusions}.

\section{Dark matter searches} \label{sec:dm}

A large number of possible extensions of the Standard Model providing viable dark matter can\-di\-dates has been proposed.
In our analysis we consider leptophilic dark matter, and specifically scenarios where dark matter couples at lowest order only to electron-positron pairs.
In fact, within leptophilic models one has the highest sensitivity when comparing to the AMS measurements of electron and positron fluxes.
Moreover, the role of electroweak radiation in these models is particularly significant, as fluxes of all stable Standard Model particles are induced,
even though the dark matter couples directly only to charged leptons. 
Given the absence of distinctive spectral features in the AMS-02 positron and electron fluxes, we obtain constraints on this class of dark matter models.

At leading order, the upper limits on the dark matter annihilation cross section we derive do not depend
on the specific choice of the model, but apply to leptophilic models in general.	For the inclusion of electroweak radiation, however, one needs to 
consider a specific model, as the model-independent fragmentation function approach~\cite{Ciafaloni:2010ti} can only be used for dark matter masses much 
larger than those of the electroweak gauge bosons, $M_{\rm DM} \gg M_{W/Z}$~\cite{AlKrPe14}. 
\begin{figure*}
  \centering
  \includegraphics[scale = 1]{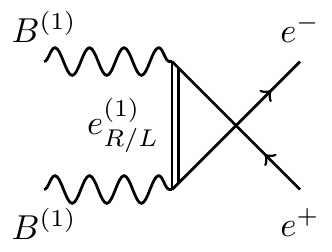}
  \includegraphics[scale = 1]{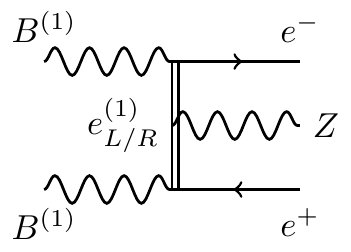}
  \includegraphics[scale = 1]{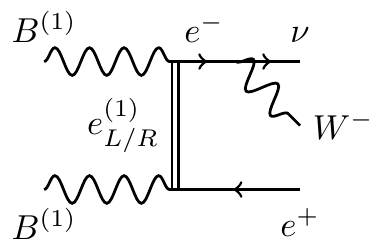}
  \includegraphics[scale = 1]{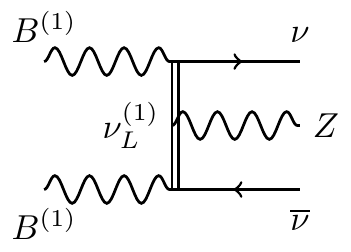}
 \caption{Contributions to Kaluza-Klein photon annihilation into an electron-positron/neutrino-antineutrino pair, including electroweak radiation.}
\label{figdiag}
\end{figure*}

To calculate the electroweak radiation we thus consider a simple 
model with a $t/u$-channel fermionic mediator and a vector dark matter candidate, as predicted in theories with Universal Extra Dimensions 
(UED)~\cite{Servant:2002aq,Kong:2005hn}.  In such UED type models, the first Kaluza-Klein excitation 
of the U(1)$_Y$ hypercharge gauge field, $B^{(1)}$, provides the dark matter candidate. 
At leading order, the annihilation process is: $B^{(1)}B^{(1)} \to e^+e^- / \overline{\nu} \nu$.
It is mediated by $t$- and $u$-channel exchange of the first Kaluza-Klein excitation of the 
electron, $e_{L,R}^{(1)}$ and of the neutrino $\nu_{L}^{(1)}$.
The electron and positron fluxes are obtained considering the primary fluxes from dark matter annihilation (in this case
the energy spectrum is simply a delta function at $E = M_{\rm{DM}}$), taking into account the particle physics evolution of the
primary decay products and propagating the particles through the Galaxy.
This model is not helicity suppressed like supersymmetric scenarios with Majorana dark matter and thus 
the induced electron and positron fluxes lead to sharp features on top of the astrophysical fluxes.
In addition, we study the effect of electroweak radiation in the calculation of the annihilation cross section
and in the generation of the fluxes.
To this end, we consider the processes $\text{DM} + \text{DM} \to e^+ + e^-$, $\text{DM} + \text{DM} \to e^+ + e^- +Z$, 
$\text{DM} + \text{DM} \to e^- + \overline{\nu} + W^+$, $\text{DM} + \text{DM} \to e^+ + \nu + W^-$ and $\text{DM} + \text{DM} \to \nu + \overline{\nu} + Z$.
\footnote{For the study of positrons and antiprotons fluxes, the channel $\text{DM} + \text{DM} \to \nu + \overline{\nu}$ is irrelevant, 
nonetheless it is necessary to preserve gauge invariance when including electroweak corrections.}
Representative Feynman diagrams for each channel are shown in Fig.~\ref{figdiag}.
To obtain the signal from dark matter annihilation at Earth
we first generate the hard process, using an in-house Monte Carlo program.
The matrix elements have been obtained with CalcHep~\cite{Belyaev:2012qa,Datta:2010us} 
and have been checked against MadGraph\_aMC@NLO~\cite{Alwall:2014hca}.
The evolution of the primary annihilation products (QED and QCD radiation, decays and hadronisation) has 
been taken into account using PYTHIA 8~\cite{SjMrSk07}. 

For the description of the propagation of cosmic ray particles, 
a Green function formalism~\cite{Cirelli:2010xx} has been used.
The Navarro-Frenk-White~(NFW)~\cite{Navarro:1995iw, Navarro:2008kc} model for the 
dark matter profile and the MED astrophysics scenario described in~\cite{Donato:2003xg,Cirelli:2010xx} 
has been used. The value for the dark matter density at the location of the Sun 
has been taken to be $\rho = 0.3$ GeV/cm$^3$~\cite{BoTr12}.

\section{Background model}\label{sec:bkg}
\begin{figure}
 \centering
 \includegraphics[scale = 0.7]{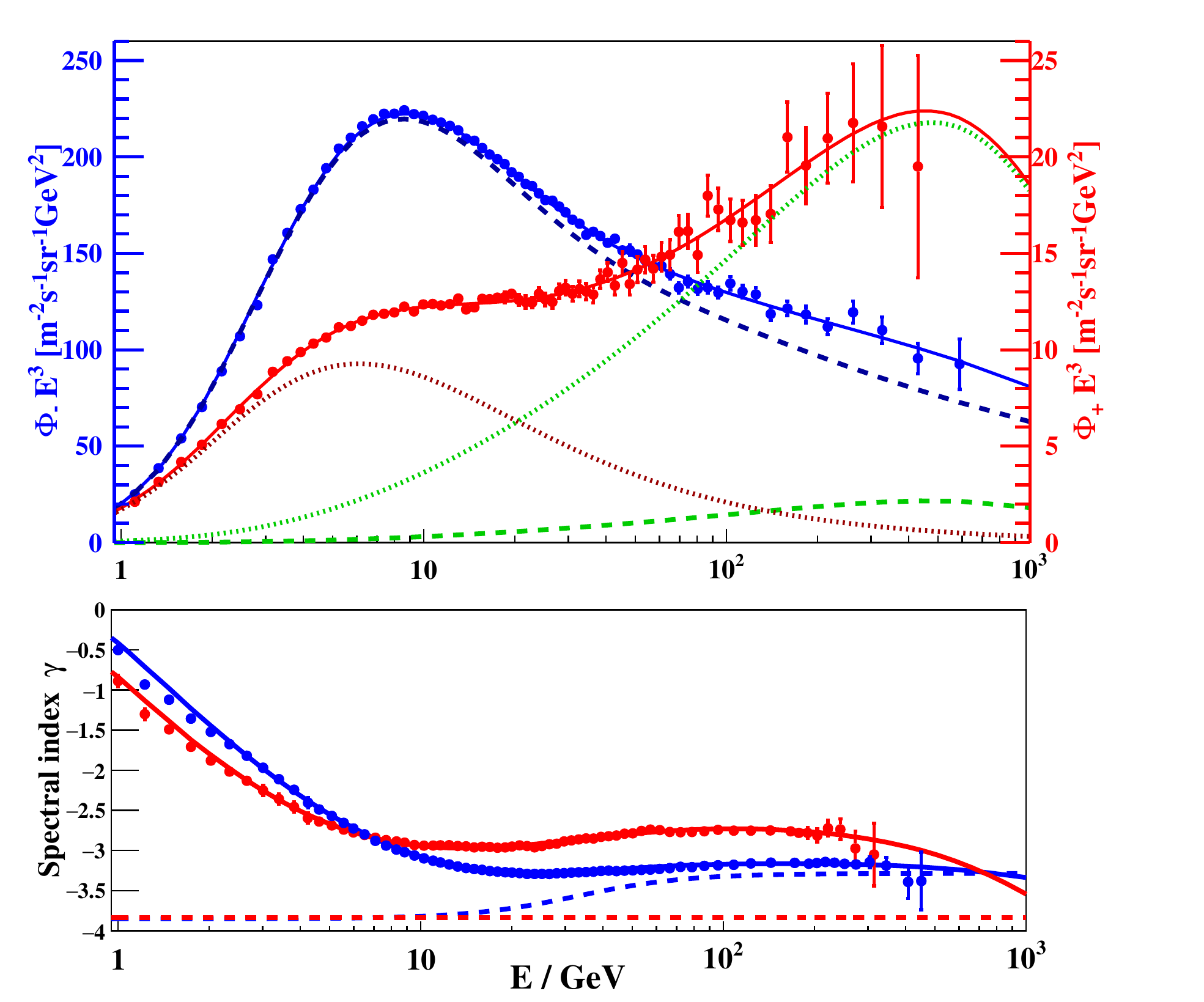}
 \caption{Top panel: electron (blue circles) and positron (red circles) fluxes measured by AMS and multiplied by $E^3$,
    as described in the text. The best fit model curve (blue solid line for positrons and red solid line for electrons)
    according to Eqs.~(\ref{eq:model}) are shown for energies above $E\geq 1\,\rm{GeV}$. 
    The separate contributions from the diffuse (dotted red and dashed blue 
    for positrons and electrons, respectively) and source term (dotted green and dashed green for positrons and 
    electrons, respectively) are also shown.
    Bottom panel: spectral index $\mathrm{d}\log\Phi/\mathrm{d}\log{}E$
    obtained from sliding-window fits to data. The solid line represents the spectral index obtained from the fit.
    The red and blue dashed lines represent the diffuse component for positrons and electrons, respectively,
    and clearly show the different behaviour of electrons and positrons.
   \label{fig:elec_posi_fit}
   }
\end{figure}
An accurate modelling of the fluxes of astrophysical origin is crucial for dark matter searches.
A successful description from first principles of the available measurements of 
electrons, positrons, protons, antiprotons and nuclei, has not been proposed yet. 
On the other hand, considering only the electron and positron data, a description in 
terms of secondary production and astrophysical sources is possible, as done for instance 
in~\cite{DiMauro:2014iia}. In the following, we search for sharp spectral features due to leading-order dark matter
annihilation into electron-positron pairs on top of such an astrophysical background that is assumed to be smooth.
For this reason, a simple phenomenological description of the background fluxes is suitable for our study.

For the description of their data on the positron fraction, the AMS
Collaboration introduced the so-called minimal phenomenological
model~\cite{AMS13}: 
\begin{subequations}
  \label{eq:min_model}
  \begin{align}
  \Phi_+(E) &= C_+ E^{-\gamma_+} + C_S E^{-\gamma_S} \exp(-E/E_S),\\
  \Phi_-(E) &= C_- E^{-\gamma_-} + C_S E^{-\gamma_S} \exp(-E/E_S),
\end{align}
\end{subequations}
and found that it describes their data extremely
well over the full energy range. In fact, the minimal model
works also for the description of the positron
flux measured by AMS, provided that the effects of solar modulation
are described in terms of the so-called force-field approximation~\cite{Gleeson:1968zza}.
However, trying to fit the electron flux with the same approach leads to 
a very poor fit with a $\chi^2/d.o.f. \sim 340/65$.
Therefore, we introduce a generalised,
phenomenological model that contains a smoothly-broken power law
for the electrons, to describe the two components
expected in the electron flux, namely secondary production and primary electrons
from astrophysical sources:
\begin{subequations}\label{eq:model}
\begin{align}
\Phi_+(E) &= (E^2/\hat{E}^2) (C_+ (\hat{E}/E_0)^{-\gamma_+} 
+ C_S (\hat{E}/E_1)^{-\gamma_S} \exp(-\lambda_S\hat{E})),  \\
 \Phi_-(E) &= (E^2/\hat{E}^2) (C_- (\hat{E}/E_0)^{-\gamma_-} (1+(\hat{E}/E_B)^{\frac{1}{b}})^{(b \Delta\gamma_-)}  
+ C_S (\hat{E}/E_1)^{-\gamma_S} \exp(-\lambda_S\hat{E})). 
\end{align}
\end{subequations}
This model contains the minimal number of parameters necessary
to obtain an accurate description of both electron and positron fluxes.
Here, $\hat{E}=E+\varphi_\pm$ is the energy of particles
in interstellar space, before interacting with the heliosphere, and
the effective potentials $\varphi_\pm$ account for 
the charge-sign dependent impact of the solar magnetic field. 
In this picture, the solar modulation potentials are the only parameters
that are expected to exhibit a time dependence.

The spectral indices
for the diffuse terms of positrons and electrons and the common source term are denoted by 
$\gamma_+, \gamma_-$ and $\gamma_S$, respectively, $E_B$ is the location of the spectral break and
$\Delta\gamma_-$ is the difference of the electrons' spectral indices before and
after the break. The smoothness of the break is described by the
parameter $b$. 
The inverse cutoff energy is given by $\lambda_S$, and the $C_\pm$ and
$C_S$ denote flux amplitudes. 
With this phenomenological model, we are able to describe electron and 
positron fluxes above 1\,GeV.

We have explicitly included the pivot
energies $E_0$ and $E_1$ in the model. They are fixed numbers that
can, in principle, be chosen at will. However, a proper choice will
substantially reduce the correlations between the model parameters in
the fit used to extract the parameters of the model 
and increase the stability of the fit. We will use
$E_0=5\,\mathrm{GeV}$ and $E_1=60\,\mathrm{GeV}$ throughout. We will
refer to the first term in the fluxes as the \emph{diffuse} term and
to the second term as the \emph{source} term, but this need not be
related to the actual physics behind the fluxes.
In this model the source term is assumed to be charge symmetric, since no
evidence for a deficit of electrons has been observed so far.
This hypothesis can be tested with the current AMS measurements
of lepton fluxes.

To determine the parameters of the model, 
we perform a $\chi^2$ minimisation using the AMS data on the separate measurements of
cosmic ray electron and positron fluxes~\cite{AMS14_2}. 
We also include the last data point of the AMS measurement of the total
$e^++e^-$ flux~\cite{AMS14_3}, covering the energy
range from 700 to 1000\,GeV, since it is statistically
independent from the other data and contains additional information
for the modelling at the highest energies.
The $\chi^2$ is obtained by adding the contributions
from these three different data sets.

The systematic uncertainties quoted by the AMS Collaboration 
vary as a function of energy in the range between $3\% - 17\%$. 
Using these systematic uncertainties the $\chi^2/n.d.f.$ of the fit is 
significantly smaller than one, showing that the systematic 
uncertainties are correlated between energy bins, as expected from the 
description of the sources of the systematic uncertainties in the corresponding publications. 
A correct treatment of these correlations would 
require the knowledge of the correlation matrix, which is not published.
In this case, the simplest assumption is that the systematic uncertainties 
consist of an uncorrelated component and a 100\% correlated component.
Therefore, for each data point, we take the published statistical uncertainty into
account, and we add an uncorrelated systematic uncertainty of only 1\,\% in
quadrature. 
We treat the remaining uncertainty with respect to the published one as 
an overall scale uncertainty on the acceptance, which effectively translates into 
an uncertainty on the normalisation of the fluxes.
A similar procedure was used by the AMS Collaboration in~\cite{AMS15}.
With this prescription we find a $\chi^2/n.d.f. \sim 1$ and an unbiased 
pull-distribution.
The fit of the positron flux alone would not allow us to constrain the model parameters
of the source term with sufficient accuracy to derive limits on a possible dark matter contribution.

The best-fit parameters and their uncertainties $\sigma_{\rm{fit}}$ are listed in Table~\ref{tab:pars}.
The $\chi^2/n.d.f.$ is $131/128$. 
The corresponding model curves are
illustrated in Fig.~\ref{fig:elec_posi_fit} for both electrons and positrons.
The same set of parameters gives good descriptions of the positron
fraction and of the $e^++e^-$ flux measured by AMS.

\begin{table}[h]
  \centering
  \begin{tabular}{lrcrcrcrl}
             parameter &        value & &  $\sigma_\mathrm{fit}$ & &  $\sigma_\mathrm{acc}$  & & $\sigma_\mathrm{scale}$ &  \\ \hline
               $C_{-}$ &        6.673 & $\,\pm\,$ &        0.183 &  $\,\pm\,$ &       0.013  & $\,\pm\,$ &        1.052  & $\mathrm{GeV}^{-1}\, \mathrm{m}^{-2}\, \mathrm{sr}^{-1}\, \mathrm{s}^{-1}$ \\
          $\gamma_{-}$ &        3.851 & $\,\pm\,$ &        0.031 &  $\,\pm\,$ &       0.007  & $\,\pm\,$ &        0.087  & \\
    $\Delta\gamma_{-}$ &        5.650 & $\,\pm\,$ &        0.561 &  $\,\pm\,$ &       0.105  & $\,\pm\,$ &        0.881  & $\times10^{-1}\, $ \\
                   $b$ &        4.171 & $\,\pm\,$ &        0.675 &  $\,\pm\,$ &       0.078  & $\,\pm\,$ &        0.466  & $\times10^{-1}\, $ \\
             $1/E_{b}$ &        3.043 & $\,\pm\,$ &        0.189 &  $\,\pm\,$ &       0.045  & $\,\pm\,$ &        0.250  & $\times10^{-2}\, \mathrm{GeV}^{-1}$ \\
               $C_{+}$ &        2.161 & $\,\pm\,$ &        0.065 &  $\,\pm\,$ &       0.014  & $\,\pm\,$ &        0.305  & $\times10^{-1}\, \mathrm{GeV}^{-1}\, \mathrm{m}^{-2}\, \mathrm{sr}^{-1}\, \mathrm{s}^{-1}$ \\
          $\gamma_{+}$ &        3.834 & $\,\pm\,$ &        0.107 &  $\,\pm\,$ &       0.007  & $\,\pm\,$ &        0.106  & \\
               $C_{S}$ &        6.189 & $\,\pm\,$ &        0.322 &  $\,\pm\,$ &       0.058  & $\,\pm\,$ &        0.494  & $\times10^{-5}\, \mathrm{GeV}^{-1}\, \mathrm{m}^{-2}\, \mathrm{sr}^{-1}\, \mathrm{s}^{-1}$ \\
          $\gamma_{S}$ &        2.525 & $\,\pm\,$ &        0.120 &  $\,\pm\,$ &       0.006  & $\,\pm\,$ &        0.045  & \\
         $\lambda_{S}$ &        1.019 & $\,\pm\,$ &        0.727 &  $\,\pm\,$ &       0.251  & $\,\pm\,$ &        0.141  & $\times10^{-3}\, \mathrm{GeV}^{-1}$ \\
         $\varphi_{-}$ &        1.406 & $\,\pm\,$ &        0.023 &  $\,\pm\,$ &       0.027  & $\,\pm\,$ &        0.096  & $\mathrm{GV}$ \\
         $\varphi_{+}$ &        1.021 & $\,\pm\,$ &        0.048 &  $\,\pm\,$ &       0.022  & $\,\pm\,$ &        0.082  & $\mathrm{GV}$ \\ \hline
  \end{tabular}
 \caption{Best-fit parameters for the model defined by
  Eqs.~(\ref{eq:model}), with parameter uncertainties due to statistical
  and uncorrelated systematic uncertainties of the data ($\sigma_\mathrm{fit}$),
  correlated systematic uncertainties ($\sigma_\mathrm{acc}$), and
  energy scale uncertainties ($\sigma_\mathrm{scale}$).  \label{tab:pars}}
\end{table}

To evaluate the scale uncertainty $\sigma_\mathrm{acc}$ introduced by the correlated
systematic uncertainties on the fit
parameters, we use the shift method~\cite{Heinrich2007}: We subtract
the 1\,\% error from the quoted systematic uncertainties in
quadrature, shift the data points upward by the remaining amount, and
repeat the fit. The same procedure is repeated for shifting the data
points downward. The value of $\sigma_\mathrm{acc}$ is then taken as
the average observed shift in the parameters from the two fits. The
resulting uncertainties (\textit{cf.}~Table~\ref{tab:pars}) 
are small compared
to the respective values of $\sigma_\mathrm{fit}$, except for those of
the solar modulation parameters $\varphi_+$ and $\varphi_-$, for which
they are of equal magnitude.
We also investigate the effect of the overall uncertainty of the
energy scale of the AMS detector on the fit results. The AMS
Collaboration quotes uncertainties of 5\,\% at 0.5\,GeV, 2\,\% in the
range from 10 to 290\,GeV, and 4\,\% at 700\,GeV~\cite{AMS14_2},
and we connect these values by straight lines in
$\log(E)$. The impact of the energy scale uncertainty on the fit
parameters can then be studied by changing the energy bin boundaries
of the data by the appropriate amount and correcting
the integral flux values accordingly. The procedure is
done for the two most extreme cases, shifting all energies upward and
downward, respectively. The corresponding uncertainty
$\sigma_\mathrm{scale}$ is calculated as the average of the
observed shifts in the parameters. It turns out that this uncertainty
is sizeable or even dominant for almost all of the fit parameters.

Finally, we tested if our model can also be used to
describe the measurements of the positron flux by
Pamela~\cite{Adriani2013} and of the electron flux by
Pamela~\cite{Adriani2011a} and Fermi-LAT~\cite{Ackermann2012}. 
We fix all parameters except $C_+$ and $C_-$.
This accounts for a possible difference in
the energy scale between the experiments, which would in the simplest case translate 
to a difference in the normalisations of the fluxes. In addition, 
since the data sets were recorded at different times, we allow 
the modulation parameters $\varphi_\pm$ to vary. 
We find in each case a $\chi^2/n.d.f. < 1$ and the best-fit values 
$C_+=(1.76\pm0.12)\times10^{-1}\, \mathrm{GeV}^{-1}\, \mathrm{m}^{-2}\,
\mathrm{sr}^{-1}\, \mathrm{s}^{-1}$ and $\varphi_+=(0.67\pm0.04)\,\mathrm{GV}$
for the Pamela positron flux, $C_-=(5.45\pm0.08)\,\mathrm{GeV}^{-1}\,
\mathrm{m}^{-2}\, \mathrm{sr}^{-1}\, \mathrm{s}^{-1}$ and
$\varphi_-=(1.11\pm0.01)\,\mathrm{GV}$ for the Pamela electron flux, and
$C_-=(6.13\pm0.21)\,\mathrm{GeV}^{-1}\, \mathrm{m}^{-2}\, \mathrm{sr}^{-1}\,
\mathrm{s}^{-1}$ for the Fermi-LAT electron flux.
Though the data points published by these
experiments are clearly inconsistent with the AMS data, 
the obtained fit parameters are within the uncertainties consistent
with those given in Table~\ref{tab:pars}, even without taking the uncertainty 
on the energy scales of PAMELA and Fermi-LAT into account.

Similar studies could be performed
using the positron fraction and the combined $e^++e^-$ flux. 
However, a precise analysis is not possible from the published
AMS results since these data sets are not statistically independent
due to an overlap in the event samples.

\section{Model independent constraints on the annihilation cross section} \label{sec:UL}
\begin{figure}
\centering
\includegraphics[scale = 0.7]{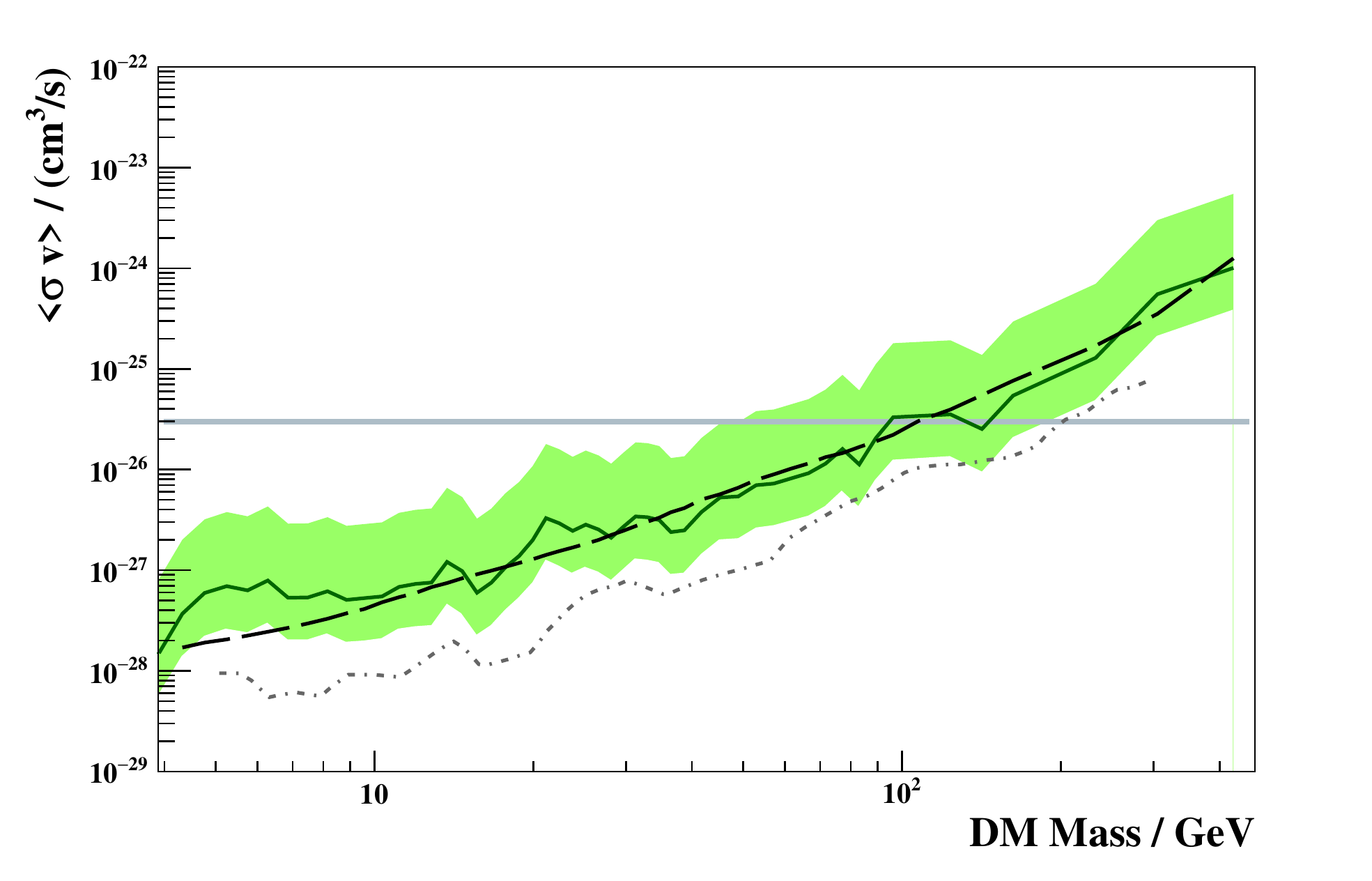}
\caption{Solid line: $95\%$ CL upper limits on the $ 2 \to 2$ annihilation cross section $\langle\sigma v\rangle$ 
for a generic model where dark matter annihilates at leading order only into electrons and positrons. 
The solid line is obtained for a specific choice of the dark matter distribution in the Galaxy (NFW profile) and
for a specific choice of the propagation parameters (MED).
Shaded band: estimate of the uncertainties due to different normalisations of the dark matter density at Earth, 
choice of the dark matter halo model and choice of the propagation model.
Black dashed line: median expected upper limits. For comparison we show the results obtained in~\cite{Bergstrom:2013jra} (grey dashed-dotted line).
Gray solid line: thermal relic cross section.
\label{fig:UL}
}
\end{figure}

For this analysis, we assume that the cosmic ray fluxes 
consist of a smooth background component that originates at high energies from some 
unspecified astrophysical source and of a sub-dominant exotic contribution 
that originates from dark matter annihilation in the Galactic halo. The latter could account 
for additional structure on top of the background predictions. 
For the description of the shape of the astrophysical background, \emph{i.e.}\ the null-hypothesis, we use Eqs.~(\ref{eq:model}).
We do not observe significant deviation from the assumed background in the measured 
fluxes. We then set constraints on leptophilic dark matter models using Wilks theorem~\cite{Wilks:1938dza}, 
namely the upper limit value on the signal normalisation is obtained increasing its value until the $\chi^2$ 
value differs by 2.71 from the null-hypotheses. The background model parameters are treated as nuisance parameters.
We first compute $95\%$ CL upper limits on the leading order $2 \to 2$ annihilation cross section.
We set upper limits on the normalisation of a possible signal due to dark matter annihilation and
we subsequently translate them into limits on the velocity averaged annihilation cross section.
These limits and the median expected upper limits obtained from $1000$ pseudo-data sets are shown in Fig.~\ref{fig:UL}. 
The pseudo-data sets are generated according to the background model, namely assuming 
that no exotic dark matter contribution is present. For each of these data sets we repeat 
the calculation of the upper limits. The median upper limits are obtained taking for each 
mass the median value of the resulting distribution.
Compared to~\cite{Bergstrom:2013jra} we find limits that are about a factor 2 weaker. 
Several aspects which contribute to this difference have been discussed already in the introduction.
In addition, the different procedure to calculate the upper limits used in~\cite{Bergstrom:2013jra} leads in
most cases to stronger limits.

We have investigated the impact of the energy scale uncertainty from the fit, of the choice of different 
cosmic ray propagation models and of the uncertainties on the antiproton production cross section. 
We find that including the uncertainties on the energy scale does not significantly affect the results for the upper limits.
More relevant are the uncertainties due to the choice of the cosmic rays propagation scenario
and dark matter halo model. To study the impact of the propagation models, we have computed the upper limit
using the MIN, MED and MAX cosmic rays propagation parameters~\cite{Cirelli:2010xx} for the NFW dark matter profile.
We also re-computed the upper limits for a fixed propagation scenario (MED) and different dark matter profiles 
(Einasto~\cite{Graham:2005xx,Navarro:2008kc}, Isothermal~\cite{Begeman:1991iy,Bahcall:1980fb},
Burkert~\cite{Burkert:1995yz,Salucci:2000ps,Gentile:2004tb,Salucci:2007tm} and Moore~\cite{Diemand:2004wh})
and different dark matter normalisations at the location of the solar system ($\rho \in [0.25, 0.7]$ GeV/cm$^3$~\cite{BoTr12}). The latter has the effect of 
trivially rescaling the upper limits curve and is the most relevant source of uncertainties.
All astrophysical parameters have been taken from~\cite{Cirelli:2010xx}.

We have recomputed the upper limits including electroweak correction but no distinguishable features are noticeable.
Indeed, multi-TeV dark matter masses can give rise to $10\%$ corrections which are sizeable for collider studies 
but are negligible with respect to astrophysical uncertainties when computing dark matter upper limits.
\section{Antiproton flux} \label{sec:pbar}

A flux of antiprotons is generated by the radiation of electroweak gauge bosons, 
$W^\pm$ and $Z$, off the primary standard model particles produced in the dark matter 
annihilation process. 
Thus, even for leptophilic dark matter models, antiprotons are produced and the 
antiproton flux can be compared to measurements to further test 
and constrain this model. For dark matter particles heavier than the 
electroweak gauge bosons, $M_{\rm DM} \gg M_{W/Z}$, the contributions 
due to electroweak radiation can be calculated in a model independent way by using 
generalised fragmentation functions~\cite{Ciafaloni:2010ti,AlKrPe14}. 
The fragmentation function approach works for models where the leading-order annihilation 
cross section is not helicity suppressed, and it provides reliable results for masses 
$M_{\rm DM} \gtrsim 5 M_{\rm W,Z} \approx 500$\,GeV~\cite{Ciafaloni:2010ti,AlKrPe14}.
However, as we are interested also in smaller dark matter masses, we consider the leptophilic dark matter model
presented earlier as a representative model.

\begin{figure}
  \centering
  \includegraphics[scale = 0.6]{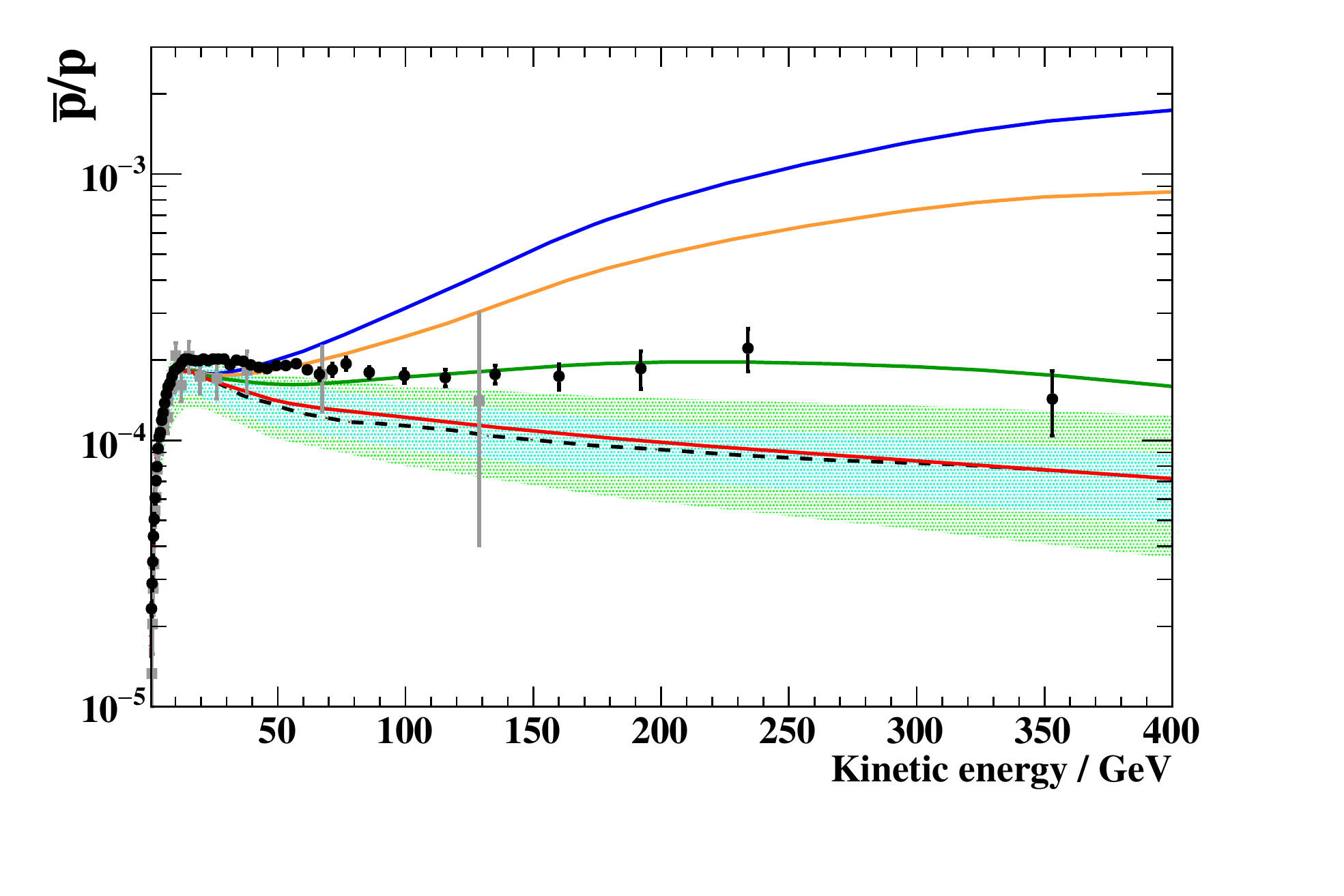}
    \caption{Black circles and grey squares: AMS-02 and PAMELA measurements, respectively.
    The solid lines represent the predictions for the antiproton-to-proton ratio for $M_{\rm{DM}} = 425, 1000, 3000, 5000$ GeV (red, green, orange and blue, respectivley). 
    The antiproton flux consists of an astrophysical component, plus an exotic component 
    due to dark matter annihilation in the Galaxy, normalised to the upper limit value of the 
    leading order annihilation cross section. The estimate of the background (dashed black line) and its uncertainties are taken from~\cite{Giesen:2015ufa}.
    More specifically, the blue band represent the uncertainties on the antiproton production cross section, while the green band represents those due to the choice of the propagation parameters.
   The dark matter signals are affected by the same sort of uncertainties, not drawn here for clarity.
   \label{fig:pbar_uncertainty}}
\end{figure}

In the previous section we have derived model independent upper limits on the $2 \to 2$ 
cross section. 
The impact on the upper limits of including electroweak radiation is found to be negligible.
However, the inclusion of electroweak radiation in our analysis is crucial as an antiproton flux 
is induced. Assuming that the dark matter annihilation cross section is at its upper limit
value, \textit{i.e.} the values represented by the black line in Fig.~\ref{fig:UL}, 
we obtain predictions for the maximum antiproton-to-proton ratio due to dark matter annihilation. 
These predictions can be compared to the measurements done by the Pamela~\cite{Adriani:2010rc} 
and AMS-02 Collaboration~\cite{AMS16}, as shown in Fig.~\ref{fig:pbar_uncertainty} 
for representative masses. 
One of the most relevant source of uncertainty is the knowledge of the cross section for
antiproton production. This has been extensively studied for instance in~\cite{Donato:2001ms,diMauro:2014zea},
according to which the uncertainties can be roughly $50\%$ outside 
the range where the antiproton productions cross section is measured.\footnote{For more details we refer to~\cite{Donato:2001ms,diMauro:2014zea} and references therein.}

The choice of the cosmic ray propagation model is a second relevant source of uncertainty, 
dominant at low energies~\cite{diMauro:2014zea}. 
In Fig.~\ref{fig:pbar_uncertainty} the background curve is the ``fiducial" antiproton-to-proton astrophysical ratio presented in~\cite{Giesen:2015ufa}. 
The uncertainties on the background are those derived in~\cite{Giesen:2015ufa}. The predictions for the antiproton-to-proton
ratio including a dark matter contribution shown in Fig.~\ref{fig:pbar_uncertainty} are affected by the same uncertainties.
For the sake of simplicity, we do not show them in the figure.

A more reliable estimate of the constraints on the dark matter model from the $\bar{p}/p$ ratio would require a systematic study of the background uncertainties 
and the correlation with the dark matter signal as presented in \textit{e.g.}~(\cite{Korsmeier:2016kha, Cuoco:2016eej,Cui:2016ppb}). We defer such a more comprehensive
analysis to a forthcoming publication.

Our analysis suggests that for dark matter masses near or above ${\cal O}(1\,{\rm TeV})$ the antiproton flux receives a sizeable contribution due 
to dark matter annihilation in the Galaxy, even for leptophilic models. It is therefore possible to
further constrain also leptophilic dark matter models 
using the complementary information contained in the antiproton flux measurements, in particular for very high dark matter masses $M_{\rm{DM}} \gtrsim 1\,{\rm TeV}$,
as shown in Fig.~\ref{fig:pbar_uncertainty}. 
The antiproton flux thus becomes relevant only for large dark matter masses.
This regime is where electroweak corrections become model independent due to the appearance of universal logarithms as shown in~\cite{Ciafaloni:2010ti, AlKrPe14}.
However, to be able to obtain robust and quantitative conclusions a better understanding of the 
astrophysical phenomena relevant to charged cosmic ray propagation models is necessary, as well as
improved measurements of the inclusive antiproton production cross section at colliders.
Additional AMS measurements of cosmic rays fluxes and ratios, like the recent boron-to-carbon ratio~\cite{AMS17}, are expected to provide new input for the modelling
of the propagation of charged cosmic rays in the Galaxy. Dedicated studies of antiproton production in proton to helium collisions
performed by the LHCb collaboration~\cite{Massacrier16, Smog}, could help to reduce the uncertainties on the antiproton production cross section. 

\section{Conclusions} \label{sec:conclusions}

We have proposed a simple phenomenological model which provides an excellent description of the 
electron and positron fluxes in cosmic rays as measured by AMS. 
Several important conclusions can be drawn from our results:
\textit{(i)} The minimal model from~\cite{AMS13} cannot be
used to derive values for the cutoff energy of its source term from
a fit to the positron fraction alone because it is too simple and does
not describe the individual fluxes.
\textit{(ii)} Neither the positron nor the electron flux shows any
sharp spectral structures. At high energies, the positron flux is
dominated by the source term while the electron flux is dominated by
the diffuse term.
\textit{(iii)} The electron flux is consistent with a charge-symmetric
source term. However, it can be shown that the electron flux alone can
equally well be described by Eqs.~(\ref{eq:model}) \emph{without}
a source term. To prove that the source term is indeed
charge-symmetric as expected from dark matter models or astrophysical
sources used to explain the observed positron excess, a solid description 
of the physical processes relevant for the diffuse term is needed.
\textit{(iv)} There is evidence for a spectral break in the electron
flux at an energy of $\sim\!30\,\mathrm{GeV}$. This might be a useful
reference point for the cross-calibration with future experiments.
\textit{(v)} The solar modulation parameters for positrons and electrons $\varphi_+$ and
$\varphi_-$ are significantly different. This shows that either the
force-field approximation breaks down in the case of cosmic ray positrons 
and/or electrons or even more additional terms are needed in the model. 

We would like to point out that the data points below 15\,GeV are
especially important to constrain the solar modulation parameters as
well as the diffuse terms. Therefore, reliable statements about source
parameters require a proper treatment of the time dependence of the
cosmic ray electron and positron fluxes. At energies below 20\,GeV, a
time dependence of the fluxes can be expected, possibly exceeding the
systematic uncertainties quoted on the average fluxes.

\textit{(vi)} Even with our assumptions on the 
degree of correlation between the systematic uncertainties of the 
AMS data points, we find very good agreement of the data with a smooth model.
On the other hand, it has been argued that a certain amount of 
spectral features (``\textit{bumpiness}'') is expected in the fluxes 
of positrons and electrons if the standard paradigm for cosmic ray 
acceleration and propagation holds, namely from the contributions 
of individual sources, and that the absence of such features would 
constitute an anomaly in our understanding of cosmic rays~\cite{Serpico}. 

We have used our improved phenomenological description of
electron and positron fluxes to place limits on the dark matter annihilation
cross section in leptophilic dark matter models. We find that 
\textit{(vii)} an appropriate description of the background is
crucial, especially for the low energy region, as most of the 
electrons and positrons produced via dark matter annihilation are soft,
since they lose energy while propagating through the Galaxy.
\textit{(viii)} Within this class of models we exclude the region
of the parameter space with  $M_{\rm{DM}} \lesssim 100 \; \rm{GeV}$ for a thermal relic,
even though this bound is somewhat diluted by the uncertainty in the normalisation
of the dark matter density. 
\textit{(ix)} The inclusion of electroweak radiation has a very small 
impact on the upper limits on the dark matter annihilation cross section. 
However, contributions due to the radiation of electroweak 
gauge boson are of crucial importance as they induce correlation
among fluxes of different particles species and, in particular, predict an antiproton
flux even within leptophilic dark matter models. This may allow to further constrain this class 
of models using measurements of the antiproton-to-proton ratio or antiproton flux. 
The comparison with Pamela and the recent AMS-02 data~\cite{AMS16} suggests
that we might be able to constrain the higher mass region ($M_{\rm{DM}} \gtrsim 3\,{\rm TeV}$) of the parameter
space, even though a careful analysis of the uncertainties is needed
in order to draw robust conclusions.

\section*{Acknowledgements}

We acknowledge support by the German Research Foundation DFG through the 
research training group ``Particle and Astroparticle Physics in the Light of the LHC", 
the Helmholtz Alliance for Astroparticle Physics (HAP) and the German Federal 
Ministry of Education and Research (BMBF).

\setlength{\bibsep}{0.15pt}
\bibliographystyle{utphys}
\bibliography{AachenPaper}

\providecommand{\href}[2]{#2}\begingroup\raggedright\begin{thebibliography}{10}

\bibitem{AMS14_2}
{\bf AMS} Collaboration, M.~Aguilar {\em et al.}, {\em {Electron and Positron
  Fluxes in Primary Cosmic Rays Measured with the Alpha Magnetic Spectrometer
  on the International Space Station}}.
\href{http://dx.doi.org/10.1103/PhysRevLett.113.121102}{Phys.Rev.Lett. {\bf
  113} (2014)  121102}.

\bibitem{Feng2014}
L.~Feng, R.-Z. Yang, H.-N. He, T.-K. Dong, Y.-Z. Fan, and J.~Chang, {\em
  {AMS-02 positron excess: new bounds on dark matter models and hint for
  primary electron spectrum hardening}}.
  \href{http://dx.doi.org/10.1016/j.physletb.2013.12.012}{Phys. Lett. {\bf
  B728} (2014)  250--255},
\href{http://arxiv.org/abs/1303.0530}{{\tt arXiv:1303.0530 [astro-ph.HE]}}.

\bibitem{Cholis:2013psa}
I.~Cholis and D.~Hooper, {\em {Dark Matter and Pulsar Origins of the Rising
  Cosmic Ray Positron Fraction in Light of New Data From AMS}}.
  \href{http://dx.doi.org/10.1103/PhysRevD.88.023013}{Phys.Rev. {\bf D88}
  (2013)  023013},
\href{http://arxiv.org/abs/1304.1840}{{\tt arXiv:1304.1840 [astro-ph.HE]}}.

\bibitem{DiMauro:2015jxa}
M.~Di~Mauro, F.~Donato, N.~Fornengo, and A.~Vittino, {\em {Dark matter vs.
  astrophysics in the interpretation of AMS-02 electron and positron data}}.
  \href{http://dx.doi.org/10.1088/1475-7516/2016/05/031}{JCAP {\bf 1605} (2016)
  no.~05, 031},
\href{http://arxiv.org/abs/1507.07001}{{\tt arXiv:1507.07001 [astro-ph.HE]}}.

\bibitem{Linden2013}
T.~Linden and S.~Profumo, {\em {Probing the Pulsar Origin of the Anomalous
  Positron Fraction with AMS-02 and Atmospheric Cherenkov Telescopes}}.
  \href{http://dx.doi.org/10.1088/0004-637X/772/1/18}{Astrophys. J. {\bf 772}
  (2013)  18},
\href{http://arxiv.org/abs/1304.1791}{{\tt arXiv:1304.1791 [astro-ph.HE]}}.

\bibitem{Kohri2016}
K.~Kohri, K.~Ioka, Y.~Fujita, and R.~Yamazaki, {\em Can we explain AMS-02
  antiproton and positron excesses simultaneously by nearby supernovae without
  pulsars or dark matter?} \href{http://dx.doi.org/10.1093/ptep/ptv193}{PTEP
  {\bf 2016} (2016) no.~2, 021E01},
\href{http://arxiv.org/abs/1505.01236}{{\tt arXiv:1505.01236 [astro-ph.HE]}}.

\bibitem{Boudaud:2014dta}
M.~Boudaud {\em et al.}, {\em {A new look at the cosmic ray positron
  fraction}}. \href{http://dx.doi.org/10.1051/0004-6361/201425197}{Astron.
  Astrophys. {\bf 575} (2015)  A67},
\href{http://arxiv.org/abs/1410.3799}{{\tt arXiv:1410.3799 [astro-ph.HE]}}.

\bibitem{CiKaRaSt08}
M.~Cirelli, M.~Kadastik, M.~Raidal, and A.~Strumia, {\em {Model-independent
  implications of the e+-, anti-proton cosmic ray spectra on properties of Dark
  Matter}}. \href{http://dx.doi.org/10.1016/j.nuclphysb.2013.05.002,
  10.1016/j.nuclphysb.2008.11.031}{Nucl.Phys. {\bf B813} (2009)  1--21},
\href{http://arxiv.org/abs/0809.2409}{{\tt arXiv:0809.2409 [hep-ph]}}.

\bibitem{DoMaBrDeSa08}
F.~Donato, D.~Maurin, P.~Brun, T.~Delahaye, and P.~Salati, {\em {Constraints on
  WIMP Dark Matter from the High Energy PAMELA $\bar{p}/p$ data}}.
  \href{http://dx.doi.org/10.1103/PhysRevLett.102.071301}{Phys.Rev.Lett. {\bf
  102} (2009)  071301},
\href{http://arxiv.org/abs/0810.5292}{{\tt arXiv:0810.5292 [astro-ph]}}.

\bibitem{Bergstrom:2009fa}
L.~Bergstr{\"o}m, J.~Edsjo, and G.~Zaharijas, {\em {Dark matter interpretation
  of recent electron and positron data}}.
  \href{http://dx.doi.org/10.1103/PhysRevLett.103.031103}{Phys.Rev.Lett. {\bf
  103} (2009)  031103},
\href{http://arxiv.org/abs/0905.0333}{{\tt arXiv:0905.0333 [astro-ph.HE]}}.

\bibitem{Finkbeiner:2010sm}
D.~P. Finkbeiner, L.~Goodenough, T.~R. Slatyer, M.~Vogelsberger, and N.~Weiner,
  {\em {Consistent Scenarios for Cosmic-Ray Excesses from Sommerfeld-Enhanced
  Dark Matter Annihilation}}.
  \href{http://dx.doi.org/10.1088/1475-7516/2011/05/002}{JCAP {\bf 1105} (2011)
   002},
\href{http://arxiv.org/abs/1011.3082}{{\tt arXiv:1011.3082 [hep-ph]}}.

\bibitem{Yuan:2013eja}
Q.~Yuan, X.-J. Bi, G.-M. Chen, Y.-Q. Guo, S.-J. Lin, {\em et al.}, {\em
  {Implications of the AMS-02 positron fraction in cosmic rays}}.
  \href{http://dx.doi.org/10.1016/j.astropartphys.2014.05.005}{Astropart.Phys.
  {\bf 60} (2015)  1--12},
\href{http://arxiv.org/abs/1304.1482}{{\tt arXiv:1304.1482 [astro-ph.HE]}}.

\bibitem{Jin:2013nta}
H.-B. Jin, Y.-L. Wu, and Y.-F. Zhou, {\em {Implications of the first AMS-02
  measurement for dark matter annihilation and decay}}.
  \href{http://dx.doi.org/10.1088/1475-7516/2013/11/026}{JCAP {\bf 1311} (2013)
   026},
\href{http://arxiv.org/abs/1304.1997}{{\tt arXiv:1304.1997 [hep-ph]}}.

\bibitem{Bertone:2008xr}
G.~Bertone, M.~Cirelli, A.~Strumia, and M.~Taoso, {\em {Gamma-ray and radio
  tests of the e+e- excess from DM annihilations}}.
  \href{http://dx.doi.org/10.1088/1475-7516/2009/03/009}{JCAP {\bf 0903} (2009)
   009},
\href{http://arxiv.org/abs/0811.3744}{{\tt arXiv:0811.3744 [astro-ph]}}.

\bibitem{Bergstrom:2013jra}
L.~Bergstr{\"o}m, T.~Bringmann, I.~Cholis, D.~Hooper, and C.~Weniger, {\em {New
  limits on dark matter annihilation from Alpha Magnetic Spectrometer Cosmic
  Ray Positron Data}}.
  \href{http://dx.doi.org/10.1103/PhysRevLett.111.171101}{Phys.Rev.Lett. {\bf
  111} (2013)  171101},
\href{http://arxiv.org/abs/1306.3983}{{\tt arXiv:1306.3983 [astro-ph.HE]}}.

\bibitem{Ibarra2014}
A.~Ibarra, A.~S. Lamperstorfer, and J.~Silk, {\em {Dark matter annihilations
  and decays after the AMS-02 positron measurements}}.
  \href{http://dx.doi.org/10.1103/PhysRevD.89.063539}{Phys. Rev. {\bf D89}
  (2014) no.~6, 063539},
\href{http://arxiv.org/abs/1309.2570}{{\tt arXiv:1309.2570 [hep-ph]}}.

\bibitem{AMS13}
{\bf AMS} Collaboration, M.~Aguilar {\em et al.}, {\em {First Result from the
  Alpha Magnetic Spectrometer on the International Space Station: Precision
  Measurement of the Positron Fraction in Primary Cosmic Rays of {0.5} -- {350}
  GeV}}.
\href{http://dx.doi.org/10.1103/PhysRevLett.110.141102}{Phys.Rev.Lett. {\bf
  110} (2013)  141102}.

\bibitem{Schael2016}
{\em {S. Schael, New results from the AMS experiment on the International Space
  Station}}, TeV Particle Astrophysics, CERN, 15 Sep 2016.

\bibitem{Ciafaloni:2010ti}
P.~Ciafaloni, D.~Comelli, A.~Riotto, F.~Sala, A.~Strumia, {\em et al.}, {\em
  {Weak Corrections are Relevant for Dark Matter Indirect Detection}}.
  \href{http://dx.doi.org/10.1088/1475-7516/2011/03/019}{JCAP {\bf 1103} (2011)
   019},
\href{http://arxiv.org/abs/1009.0224}{{\tt arXiv:1009.0224 [hep-ph]}}.

\bibitem{AlKrPe14}
L.~Ali~Cavasonza, M.~Kr{\"a}mer, and M.~Pellen, {\em {Electroweak fragmentation
  functions for dark matter annihilation}}.
  \href{http://dx.doi.org/10.1088/1475-7516/2015/02/021}{JCAP {\bf 1502} (2015)
  no.~02, 021},
\href{http://arxiv.org/abs/1409.8226}{{\tt arXiv:1409.8226 [hep-ph]}}.

\bibitem{Servant:2002aq}
G.~Servant and T.~M. Tait, {\em {Is the lightest Kaluza-Klein particle a viable
  dark matter candidate?}}
  \href{http://dx.doi.org/10.1016/S0550-3213(02)01012-X}{Nucl.Phys. {\bf B650}
  (2003)  391--419},
\href{http://arxiv.org/abs/hep-ph/0206071}{{\tt arXiv:hep-ph/0206071
  [hep-ph]}}.

\bibitem{Kong:2005hn}
K.~Kong and K.~T. Matchev, {\em {Precise calculation of the relic density of
  Kaluza-Klein dark matter in universal extra dimensions}}.
  \href{http://dx.doi.org/10.1088/1126-6708/2006/01/038}{JHEP {\bf 0601} (2006)
   038},
\href{http://arxiv.org/abs/hep-ph/0509119}{{\tt arXiv:hep-ph/0509119
  [hep-ph]}}.

\bibitem{Belyaev:2012qa}
A.~Belyaev, N.~D. Christensen, and A.~Pukhov, {\em {CalcHEP 3.4 for collider
  physics within and beyond the Standard Model}}.
  \href{http://dx.doi.org/10.1016/j.cpc.2013.01.014}{Comput.Phys.Commun. {\bf
  184} (2013)  1729--1769},
\href{http://arxiv.org/abs/1207.6082}{{\tt arXiv:1207.6082 [hep-ph]}}.

\bibitem{Datta:2010us}
A.~Datta, K.~Kong, and K.~T. Matchev, {\em {Minimal Universal Extra Dimensions
  in CalcHEP/CompHEP}}.
  \href{http://dx.doi.org/10.1088/1367-2630/12/7/075017}{New J.Phys. {\bf 12}
  (2010)  075017},
\href{http://arxiv.org/abs/1002.4624}{{\tt arXiv:1002.4624 [hep-ph]}}.

\bibitem{Alwall:2014hca}
J.~Alwall, R.~Frederix, S.~Frixione, V.~Hirschi, F.~Maltoni, {\em et al.}, {\em
  {The automated computation of tree-level and next-to-leading order
  differential cross sections, and their matching to parton shower
  simulations}}. \href{http://dx.doi.org/10.1007/JHEP07(2014)079}{JHEP {\bf
  1407} (2014)  079},
\href{http://arxiv.org/abs/1405.0301}{{\tt arXiv:1405.0301 [hep-ph]}}.

\bibitem{SjMrSk07}
T.~Sjostrand, S.~Mrenna, and P.~Z. Skands, {\em {A Brief Introduction to PYTHIA
  8.1}}. \href{http://dx.doi.org/10.1016/j.cpc.2008.01.036}{Comput.Phys.Commun.
  {\bf 178} (2008)  852--867},
\href{http://arxiv.org/abs/0710.3820}{{\tt arXiv:0710.3820 [hep-ph]}}.

\bibitem{Cirelli:2010xx}
M.~Cirelli, G.~Corcella, A.~Hektor, G.~Hutsi, M.~Kadastik, {\em et al.}, {\em
  {PPPC 4 DM ID: A Poor Particle Physicist Cookbook for Dark Matter Indirect
  Detection}}. \href{http://dx.doi.org/10.1088/1475-7516/2012/10/E01,
  10.1088/1475-7516/2011/03/051}{JCAP {\bf 1103} (2011)  051},
\href{http://arxiv.org/abs/1012.4515}{{\tt arXiv:1012.4515 [hep-ph]}}.

\bibitem{Navarro:1995iw}
J.~F. Navarro, C.~S. Frenk, and S.~D. White, {\em {The Structure of cold dark
  matter halos}}. \href{http://dx.doi.org/10.1086/177173}{Astrophys.J. {\bf
  462} (1996)  563--575},
\href{http://arxiv.org/abs/astro-ph/9508025}{{\tt arXiv:astro-ph/9508025
  [astro-ph]}}.

\bibitem{Navarro:2008kc}
J.~F. Navarro, A.~Ludlow, V.~Springel, J.~Wang, M.~Vogelsberger, {\em et al.},
  {\em {The Diversity and Similarity of Cold Dark Matter Halos}}.
\href{http://arxiv.org/abs/0810.1522}{{\tt arXiv:0810.1522 [astro-ph]}}.

\bibitem{Donato:2003xg}
F.~Donato, N.~Fornengo, D.~Maurin, and P.~Salati, {\em Antiprotons in cosmic
  rays from neutralino annihilation}.
  \href{http://dx.doi.org/10.1103/PhysRevD.69.063501}{Phys. Rev. {\bf D69}
  (2004)  063501},
\href{http://arxiv.org/abs/astro-ph/0306207}{{\tt arXiv:astro-ph/0306207
  [astro-ph]}}.

\bibitem{BoTr12}
J.~Bovy and S.~Tremaine, {\em {On the local dark matter density}}.
  \href{http://dx.doi.org/10.1088/0004-637X/756/1/89}{Astrophys.J. {\bf 756}
  (2012)  89},
\href{http://arxiv.org/abs/1205.4033}{{\tt arXiv:1205.4033 [astro-ph.GA]}}.

\bibitem{DiMauro:2014iia}
M.~Di~Mauro, F.~Donato, N.~Fornengo, R.~Lineros, and A.~Vittino, {\em
  {Interpretation of AMS-02 electrons and positrons data}}.
  \href{http://dx.doi.org/10.1088/1475-7516/2014/04/006}{JCAP {\bf 1404} (2014)
   006},
\href{http://arxiv.org/abs/1402.0321}{{\tt arXiv:1402.0321 [astro-ph.HE]}}.

\bibitem{Gleeson:1968zza}
L.~Gleeson and W.~Axford, {\em {Solar Modulation of Galactic Cosmic Rays}}.
\href{http://dx.doi.org/10.1086/149822}{Astrophys.J. {\bf 154} (1968)  1011}.

\bibitem{AMS14_3}
{\bf AMS} Collaboration, M.~Aguilar {\em et al.}, {\em {Precision Measurement
  of the ($e^+ + e^-ˆ'$) Flux in Primary Cosmic Rays from 0.5 GeV to 1 TeV with
  the Alpha Magnetic Spectrometer on the International Space Station}}.
\href{http://dx.doi.org/10.1103/PhysRevLett.113.221102}{Phys.Rev.Lett. {\bf
  113} (2014)  221102}.

\bibitem{AMS15}
{\bf AMS} Collaboration, M.~Aguilar {\em et al.}, {\em {Precision Measurement
  of the Proton Flux in Primary Cosmic Rays from Rigidity 1 GV to 1.8 TV with
  the Alpha Magnetic Spectrometer on the International Space Station}}.
\href{http://dx.doi.org/10.1103/PhysRevLett.114.171103}{Phys.Rev.Lett. {\bf
  114} (2015) no.~17, 171103}.

\bibitem{Heinrich2007}
J.~Heinrich and L.~Lyons, {\em {Systematic errors}}.
\href{http://dx.doi.org/10.1146/annurev.nucl.57.090506.123052}{Ann. Rev. Nucl.
  Part. Sci. {\bf 57} (2007)  145--169}.

\bibitem{Adriani2013}
{\bf PAMELA} Collaboration, O.~Adriani {\em et al.}, {\em {Cosmic-Ray Positron
  Energy Spectrum Measured by PAMELA}}.
  \href{http://dx.doi.org/10.1103/PhysRevLett.111.081102}{Phys. Rev. Lett. {\bf
  111} (2013)  081102},
\href{http://arxiv.org/abs/1308.0133}{{\tt arXiv:1308.0133 [astro-ph.HE]}}.

\bibitem{Adriani2011a}
{\bf PAMELA} Collaboration, O.~Adriani {\em et al.}, {\em {The cosmic-ray
  electron flux measured by the PAMELA experiment between 1 and 625 GeV}}.
  \href{http://dx.doi.org/10.1103/PhysRevLett.106.201101}{Phys. Rev. Lett. {\bf
  106} (2011)  201101},
\href{http://arxiv.org/abs/1103.2880}{{\tt arXiv:1103.2880 [astro-ph.HE]}}.

\bibitem{Ackermann2012}
{\bf Fermi-LAT} Collaboration, M.~Ackermann {\em et al.}, {\em {Measurement of
  separate cosmic-ray electron and positron spectra with the Fermi Large Area
  Telescope}}. \href{http://dx.doi.org/10.1103/PhysRevLett.108.011103}{Phys.
  Rev. Lett. {\bf 108} (2012)  011103},
\href{http://arxiv.org/abs/1109.0521}{{\tt arXiv:1109.0521 [astro-ph.HE]}}.

\bibitem{Wilks:1938dza}
S.~S. Wilks, {\em {The Large-Sample Distribution of the Likelihood Ratio for
  Testing Composite Hypotheses}}.
\href{http://dx.doi.org/10.1214/aoms/1177732360}{Annals Math. Statist. {\bf 9}
  (1938) no.~1, 60--62}.

\bibitem{Graham:2005xx}
A.~W. Graham, D.~Merritt, B.~Moore, J.~Diemand, and B.~Terzic, {\em {Empirical
  models for Dark Matter Halos. I. Nonparametric Construction of Density
  Profiles and Comparison with Parametric Models}}.
  \href{http://dx.doi.org/10.1086/508988}{Astron.J. {\bf 132} (2006)
  2685--2700},
\href{http://arxiv.org/abs/astro-ph/0509417}{{\tt arXiv:astro-ph/0509417
  [astro-ph]}}.

\bibitem{Begeman:1991iy}
K.~Begeman, A.~Broeils, and R.~Sanders, {\em {Extended rotation curves of
  spiral galaxies: Dark haloes and modified dynamics}}.
Mon.Not.Roy.Astron.Soc. {\bf 249} (1991)  523.

\bibitem{Bahcall:1980fb}
J.~N. Bahcall and R.~Soneira, {\em {The Universe at faint magnetidues. 2.
  Models for the predicted star counts}}.
\href{http://dx.doi.org/10.1086/190685}{Astrophys.J.Suppl. {\bf 44} (1980)
  73--110}.

\bibitem{Burkert:1995yz}
A.~Burkert, {\em {The Structure of dark matter halos in dwarf galaxies}}.
  \href{http://dx.doi.org/10.1086/309560}{IAU Symp. {\bf 171} (1996)  175},
\href{http://arxiv.org/abs/astro-ph/9504041}{{\tt arXiv:astro-ph/9504041
  [astro-ph]}}.

\bibitem{Salucci:2000ps}
P.~Salucci and A.~Burkert, {\em {Dark matter scaling relations}}.
  \href{http://dx.doi.org/10.1086/312747}{Astrophys.J. {\bf 537} (2000)
  L9--L12},
\href{http://arxiv.org/abs/astro-ph/0004397}{{\tt arXiv:astro-ph/0004397
  [astro-ph]}}.

\bibitem{Gentile:2004tb}
G.~Gentile, P.~Salucci, U.~Klein, D.~Vergani, and P.~Kalberla, {\em {The Cored
  distribution of dark matter in spiral galaxies}}.
  \href{http://dx.doi.org/10.1111/j.1365-2966.2004.07836.x}{Mon.Not.Roy.Astron.Soc.
  {\bf 351} (2004)  903},
\href{http://arxiv.org/abs/astro-ph/0403154}{{\tt arXiv:astro-ph/0403154
  [astro-ph]}}.

\bibitem{Salucci:2007tm}
P.~Salucci, A.~Lapi, C.~Tonini, G.~Gentile, I.~Yegorova, {\em et al.}, {\em
  {The Universal Rotation Curve of Spiral Galaxies. 2. The Dark Matter
  Distribution out to the Virial Radius}}.
  \href{http://dx.doi.org/10.1111/j.1365-2966.2007.11696.x}{Mon.Not.Roy.Astron.Soc.
  {\bf 378} (2007)  41--47},
\href{http://arxiv.org/abs/astro-ph/0703115}{{\tt arXiv:astro-ph/0703115
  [ASTRO-PH]}}.

\bibitem{Diemand:2004wh}
J.~Diemand, B.~Moore, and J.~Stadel, {\em {Convergence and scatter of cluster
  density profiles}}.
  \href{http://dx.doi.org/10.1111/j.1365-2966.2004.08094.x}{Mon.Not.Roy.Astron.Soc.
  {\bf 353} (2004)  624},
\href{http://arxiv.org/abs/astro-ph/0402267}{{\tt arXiv:astro-ph/0402267
  [astro-ph]}}.

\bibitem{Adriani:2010rc}
{\bf PAMELA} Collaboration, O.~Adriani {\em et al.}, {\em {PAMELA results on
  the cosmic-ray antiproton flux from 60 MeV to 180 GeV in kinetic energy}}.
  \href{http://dx.doi.org/10.1103/PhysRevLett.105.121101}{Phys.Rev.Lett. {\bf
  105} (2010)  121101},
\href{http://arxiv.org/abs/1007.0821}{{\tt arXiv:1007.0821 [astro-ph.HE]}}.

\bibitem{AMS16}
{\bf AMS} Collaboration, M.~Aguilar {\em et al.}, {\em {Antiproton Flux,
  Antiproton-to-Proton Flux Ratio, and Properties of Elementary Particle Fluxes
  in Primary Cosmic Rays Measured with the Alpha Magnetic Spectrometer on the
  International Space Station}}.
\href{http://dx.doi.org/10.1103/PhysRevLett.117.091103}{Phys. Rev. Lett. {\bf
  117} (2016) no.~9, }.

\bibitem{Donato:2001ms}
F.~Donato, D.~Maurin, P.~Salati, A.~Barrau, G.~Boudoul, and R.~Taillet, {\em
  {Anti-protons from spallations of cosmic rays on interstellar matter}}.
  \href{http://dx.doi.org/10.1086/323684}{Astrophys. J. {\bf 563} (2001)
  172--184},
\href{http://arxiv.org/abs/astro-ph/0103150}{{\tt arXiv:astro-ph/0103150
  [astro-ph]}}.

\bibitem{diMauro:2014zea}
M.~di~Mauro, F.~Donato, A.~Goudelis, and P.~D. Serpico, {\em {New evaluation of
  the antiproton production cross section for cosmic ray studies}}.
  \href{http://dx.doi.org/10.1103/PhysRevD.90.085017}{Phys. Rev. {\bf D90}
  (2014) no.~8, 085017},
\href{http://arxiv.org/abs/1408.0288}{{\tt arXiv:1408.0288 [hep-ph]}}.

\bibitem{Giesen:2015ufa}
G.~Giesen, M.~Boudaud, Y.~Génolini, V.~Poulin, M.~Cirelli, P.~Salati, and P.~D.
  Serpico, {\em {AMS-02 antiprotons, at last! Secondary astrophysical component
  and immediate implications for Dark Matter}}.
  \href{http://dx.doi.org/10.1088/1475-7516/2015/09/023,
  10.1088/1475-7516/2015/9/023}{JCAP {\bf 1509} (2015) no.~09, 023},
\href{http://arxiv.org/abs/1504.04276}{{\tt arXiv:1504.04276 [astro-ph.HE]}}.

\bibitem{Korsmeier:2016kha}
M.~Korsmeier and A.~Cuoco, {\em {Galactic cosmic-ray propagation in the light
  of AMS-02: Analysis of protons, helium, and antiprotons}}.
  \href{http://dx.doi.org/10.1103/PhysRevD.94.123019}{Phys. Rev. {\bf D94}
  (2016) no.~12, 123019},
\href{http://arxiv.org/abs/1607.06093}{{\tt arXiv:1607.06093 [astro-ph.HE]}}.

\bibitem{Cuoco:2016eej}
A.~Cuoco, M.~Kr{\"a}mer, and M.~Korsmeier, {\em {Novel dark matter constraints
  from antiprotons in the light of AMS-02}}.
\href{http://arxiv.org/abs/1610.03071}{{\tt arXiv:1610.03071 [astro-ph.HE]}}.

\bibitem{Cui:2016ppb}
M.-Y. Cui, Q.~Yuan, Y.-L.~S. Tsai, and Y.-Z. Fan, {\em {A possible dark matter
  annihilation signal in the AMS-02 antiproton data}}.
\href{http://arxiv.org/abs/1610.03840}{{\tt arXiv:1610.03840 [astro-ph.HE]}}.

\bibitem{AMS17}
{\bf AMS} Collaboration, M.~Aguilar {\em et al.}, {\em {Precision Measurement
  of the Boron to Carbon Flux Ratio in Cosmic Rays from 1.9 GV to 2.6 TV with
  the Alpha Magnetic Spectrometer on the International Space Station}}.
\href{http://dx.doi.org/10.1103/PhysRevLett.117.231102}{Phys. Rev. Lett. {\bf
  117} (2016) no.~23, 231102}.

\bibitem{Massacrier16}
{\em {L. Massacrier}}, CERN Seminar, 28 June 2016.

\bibitem{Smog}
{\em {SMOG as a Fixed target in the LHC}},
  https://lbtwiki.cern.ch/bin/view/velo/smogasfixedtarget.

\bibitem{Serpico}
P.~Serpico, {\em Possible physics scenarios behind cosmic-ray anomalies}. 34th
  International Cosmic Ray Conference (ICRC), The Hague, The Netherlands (2015)
   .

\end{thebibliography}\endgroup


\end{document}